\documentclass[journal]{IEEEtran}
\ifCLASSINFOpdf
\else
\fi
\usepackage{amsfonts,amsmath,amssymb}
\usepackage{mathtools}
\usepackage{mdwmath}
\usepackage{cuted}
\usepackage{mdwtab}
\usepackage{amsthm}           % for designing theorems
\usepackage{bm}               % provides bold math \bm{} command (vectors)
\usepackage{units}            % for nicely typset units \unit[10]{kHz}
\usepackage{graphicx}
\usepackage{epstopdf}
\usepackage[linesnumbered,ruled,vlined]{algorithm2e}

\usepackage{url}
\usepackage{paralist}
\usepackage{color}
\usepackage{authblk}
\usepackage{cleveref}
\usepackage{float}
\usepackage[caption = false]{subfig}
\newcommand{\Fig}[1]{Fig.~\ref{fig:#1}}

\newcommand{\Eq}[1]{Eq.~(\ref{eq:#1})}

\newcommand{\Lc}{\mathcal{L}}

\newcommand{\Rs}{\mathbb{R}^2}

\newcommand{\Ed}{\mathbb{E}}

\newcommand{\Pd}{\mathbb{P}}
\newcommand{\sinr}{\mathrm{SINR}}

\newcommand{\dr}{\mathrm{d}}

\newcommand{\Ru}{v}

\newcommand{\los}{\texttt{l}}
\newcommand{\nlos}{\texttt{n}}


\begin{document}
\title{Impact of UAV Antenna Configuration on Wireless Connectivity in Urban Environments}

\author{Boris Galkin,
        Jacek~Kibi\l{}da,
        and~Luiz~A. DaSilva
%\thanks{This material is based upon works supported by the Science Foundation
%Ireland under Grants No. 10/IN.1/I3007 and 14/US/I3110. B. Galkin, J. Kibi\l{}da, and L. DaSilva are with %CONNECT, Trinity College Dublin, Ireland, email:  \{galkinb,kibildj,dasilval\}@tcd.ie.}% <-this % stops a space
}

\affil{CONNECT, Trinity College Dublin, Ireland \\
{email: \{galkinb,kibildj,dasilval\}@tcd.ie}}

\maketitle

\begin{abstract}
The growing presence of UAVs has led operators to explore the issue of how to provide wireless service to UAVs. The achieveable service quality will be affected by a number of factors, including the type of antenna that the UAVs are equipped with and whether to use existing cellular infrastructure or deploy dedicated ground stations to service these new user devices. In this paper we employ stochastic geometry to carry out an analysis of the UAV coverage probability that can be achieved with a network of dedicated ground stations using a sub-6GHz technology such as LTE under different deployment conditions, given different types of UAV antennas. We provide analytical expressions for the coverage probability; we also evaluate the performance of the GS network against the case where the UAVs are served by a typical terrestrial BS network.
\end{abstract}

\begin{IEEEkeywords}
UAV networks, wireless backhaul, Poisson point process, stochastic geometry
\end{IEEEkeywords}

% \fi
\IEEEpeerreviewmaketitle

\vspace{-5mm}
\section{Introduction}
\vspace{-2mm}
%The growing popularity of unmanned aerial vehicles (UAVs) for use in commercial and public safety applications has highlighted the need for fast and reliable UAV wireless connectivity. Due to their airborne positions UAVs experience very different channel propagation conditions than terrestrial devices; this has resulted in a wide body of work on measuring and modelling the wireless UAV channel for a variety of environments and wireless technologies \cite{Khawaja_2018}. Most recently, the 3GPP has released a specification on the performance of terrestrial LTE networks when serving low-altitude UAVs, including a channel propagation model to be used for UAV network simulations \cite{3GPP_2018}. The 3GPP results suggest that terrestrial base stations (BSs) are capable of providing low-altitude UAVs with reliable wireless channels for command \& control (C\&C) signalling, provided the UAVs are equipped with steerable directional antennas for interference mitigation. The 3GPP results corroborate other recent investigations of the performance of terrestrial networks serving aerial users \cite{Lin_2017}, \cite{MahdiAzari_20172}.

Due to their ability to intelligently move in three-dimensional space, unmanned aerial vehicles (UAVs) have attracted the attention of the wireless community for use in wireless sensor networks \cite{Heimfarth_2014}, public safety networks \cite{Merwaday_2015} and as flying infrastructure providing service to demand hotspots in densely populated areas \cite{BorYaliniz_20162}. Whereas terrestrial infrastructure can make use of wired backhauls into the core network, UAV-mounted infrastructure
%, due to its airborne nature, 
must rely on a wireless backhaul link. This wireless link may be based on free space optical (FSO) communication \cite{Alzenad_2018}, millimeter wave \cite{Xiao_2016}, or it may use a sub-6GHz technology such as LTE. A number of works have been published on the use of existing terrestrial BS networks for providing wireless connectivity to low-altitude UAVs. In \cite{Rohde_2012} the authors simulate a network outage scenario where UAVs replace offline BSs, with the UAVs wirelessly backhauling into adjacent, functioning BSs. In \cite{Lin_2017} the authors simulate an LTE network in a rural environment and demonstrate how it can serve UAVs operating at heights below 120m. In \cite{MahdiAzari_20172} the authors consider the performance of UAV-BS links in different environments, given down-tilted BS antennas and omnidirectional UAV antennas. In \cite{Geraci_2018} the authors explore the use of massive MIMO in terrestrial BSs for the UAV backhaul. 

The general consensus to date is that terrestrial BS networks are capable of serving low-altitude UAV users. However, due to their downtilted antennas, which are not designed for aerial users, they may be unsuitable for certain types of UAV applications. As the number of UAVs that require high quality service increases, network operators may need to deploy dedicated ground stations (GSs), using dedicated infrastructure and spectral resources, to provide wireless connectivity to these UAVs. In our previous work \cite{Galkin_2018} we explored the performance of networks of dedicated GSs, using LTE and millimeter wave technologies, and demonstrated how even at low densities the networks can provide UAVs with high data rate wireless channels.

One of the key questions that has not been adequately explored in the existing state of the art is the impact of UAV antenna configuration on their connectivity to ground stations. Given the wide range of UAV applications and the growing number of UAVs in the market we may expect a variety of UAV antenna configurations of differing complexity and efficiency to be present in a network. The variation in antenna design will radically affect the resulting network performance, which will, in turn, dictate how network operators approach the problem of providing wireless service to UAVs. In this letter we extend our work in \cite{Galkin_2018} to explore the achieveable performance of a dedicated GS network for different UAV antenna types, namely omnidirectional, fixed directional, and steerable directional. One of our contributions is to provide a stochastic geometry model which is general enough to capture the performance impact of these antenna types. By comparing the network behaviour for different UAV antenna types we are able to demonstrate the exact impact that UAV antenna directionality combined with intelligent beam alignment can have on network performance. Additionally, we compare the numerical results of our model against simulations of UAV service from terrestrial BS networks, as envisioned in the state of the art. This comparison allows us to quantify the benefits of dedicated GS networks against the existing terrestrial BS networks for UAV service. The numerical results also allow us to demonstrate how the UAV antenna configuration and their height above ground will be a major factor in determing whether an operator must deploy dedicated infrastructure for connecting UAVs, or if they can rely on the existing terrestrial BS network.

\vspace{-5mm}
\section{System Model}

\begin{figure}[t!]
\vspace{-5mm}
\centering
	\subfloat{\includegraphics[width=.45\textwidth]{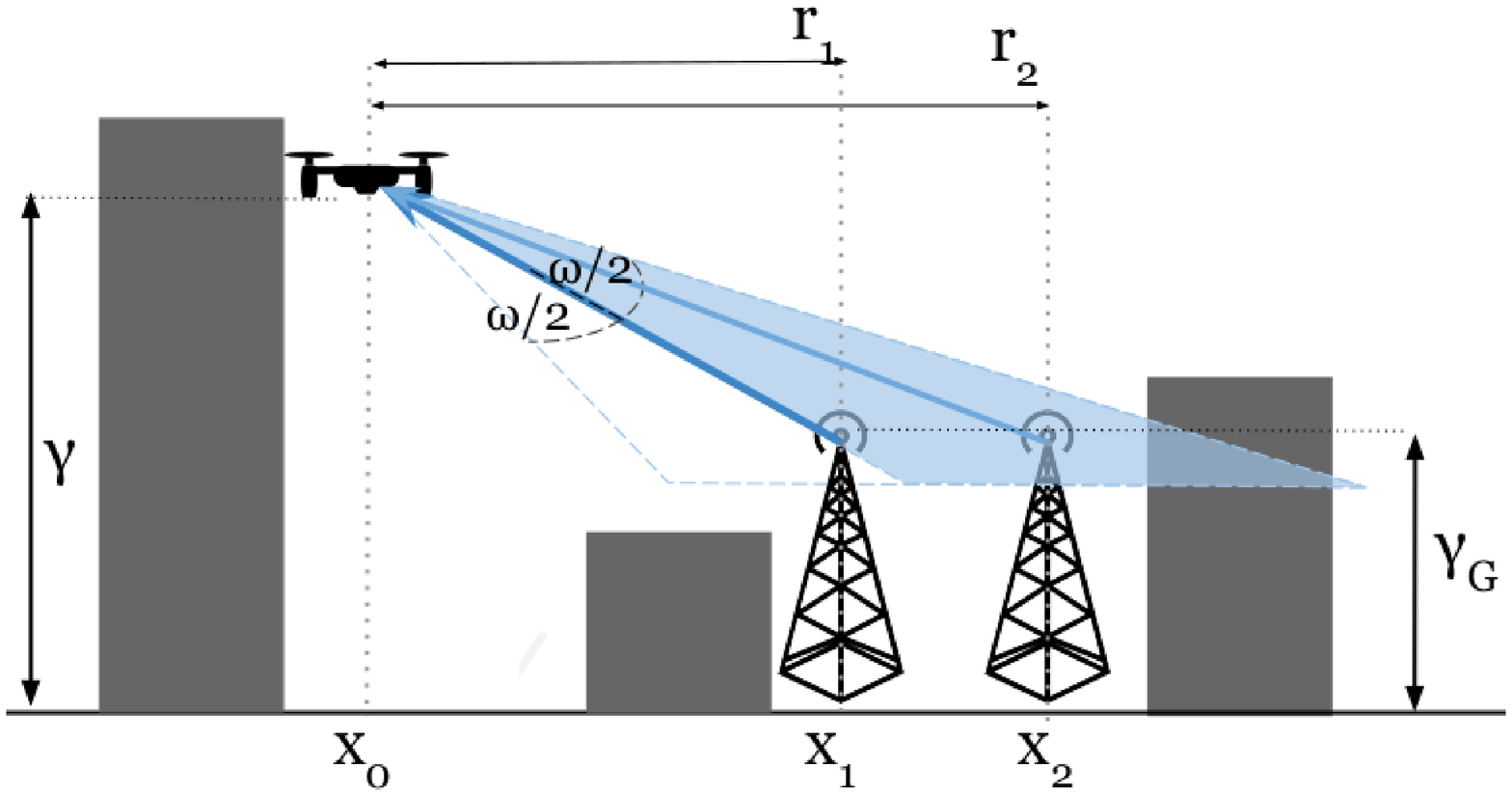}}\\
%	\vspace{-13mm}
	\subfloat{\includegraphics[width=.45\textwidth]{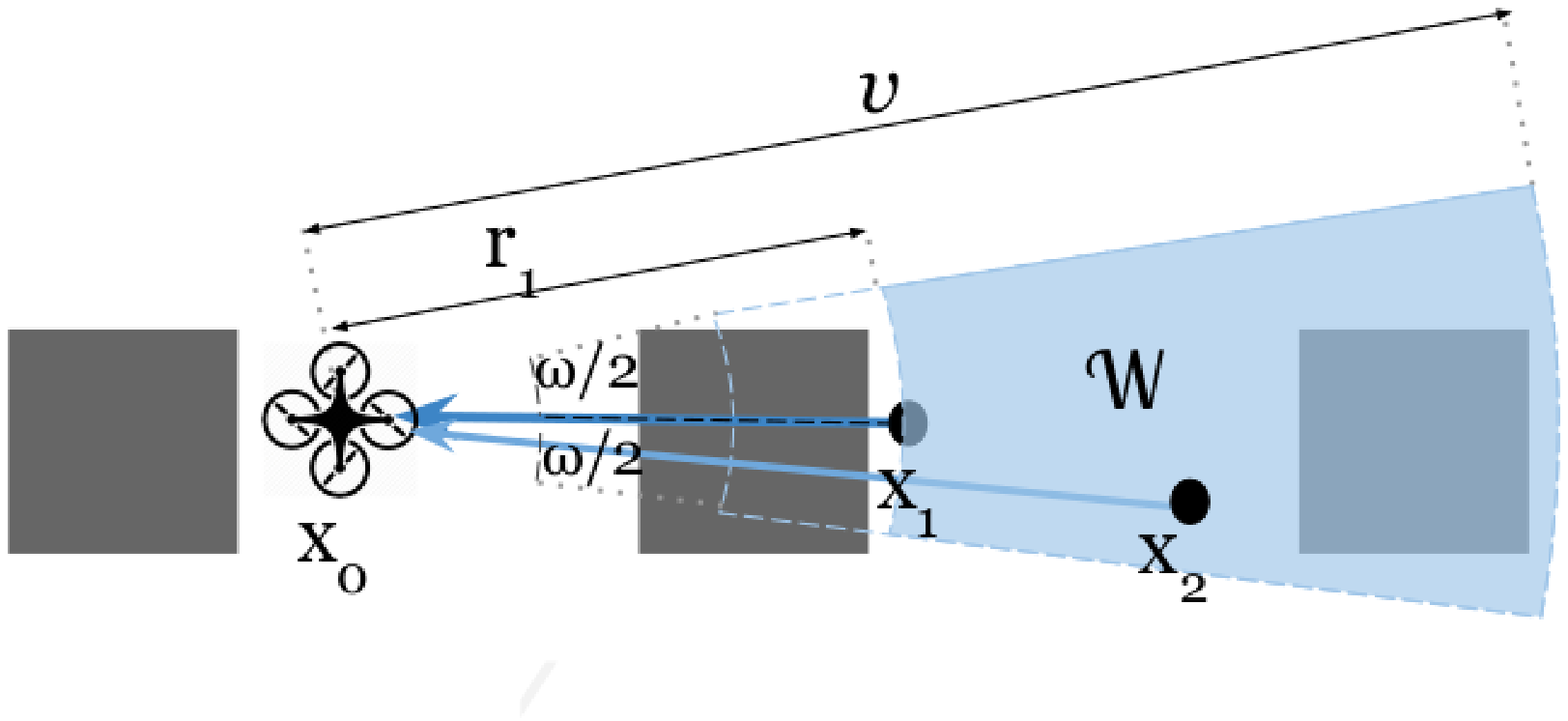}}
	\vspace{-5mm}
	\caption{
	Side and top view showing a UAV in an urban environment at a height $\gamma$, positioned above $x_0$ with antenna beamwidth $\omega$. The UAV establishes a link with its nearest GS at $x_1$ and aligns its antenna main lobe on the GS location; the blue area $\mathcal{W}$ illuminated by the main lobe denotes the region where interferers may be found. The GS at $x_2$ falls inside this area and produces interference. 
	\vspace{-5mm}
	}
	\label{fig:drone_network}
\end{figure}
\vspace{-3mm}
In this section we set up a system model of a network of GSs providing wireless connectivity to UAVs in the air. We model the network of GSs as a Poisson point process (PPP) $\Phi = \{x_1 , x_2 , ...\} \subset \Rs$ of intensity $\lambda$, where elements $x_i\in \Rs$ represent the GS locations.  The GSs have a height $\gamma_{G}$ above ground, as depicted in \Fig{drone_network}. We consider a single reference UAV, positioned above the origin $x_0 = (0,0)$ at a height $\gamma$. This reference UAV represents a typical UAV belonging to a UAV network with an arbitrary spatial distribution. Let $r_i = ||x_i||$ denote the horizontal distance between the GS $i$ and the reference UAV, and let $\phi_i = \arctan(\Delta \gamma/r_i)$ denote the vertical angle, where $\Delta \gamma = \gamma - \gamma_{G}$. 

We consider three different types of antennas that the UAV may use to communicate with its GS network.
\begin{enumerate} 
\item An omnidirectional antenna, with antenna gain $\eta = 1$. 
\item A directional antenna tilted down to give a cone-shaped radiation lobe directly beneath the UAV \cite{Galkin_2017}. Using the approximations (2-26) and (2-49) in \cite{Balanis_2005} and assuming perfect radiation efficiency, the antenna gain can be expressed as $\eta = 16\pi/(\omega^2)$ inside the main lobe and $\eta=0$ outside, where $\omega$ denotes the antenna beamwidth. 
\item A directional antenna which the UAV can intelligently steer and align with its serving GS, as depicted in \Fig{drone_network}. The antenna has a horizontal and vertical beamwidth $\omega$ and a rectangular radiation pattern; the antenna gain is given as $\eta = 16\pi/(\omega^2)$ inside the main lobe and $\eta=0$ outside. 
\end{enumerate}

The UAV selects the nearest GS as its serving GS\footnote{Given a low density of GSs with antennas tilted towards the sky the strongest received signal power at the UAV is expected to come from the nearest GS.}; we denote this GS as $x_1$ and its distance to the UAV as $r_1$. the UAV antenna reception pattern illuminates an area beyond $x_1$ which we denote as $\mathcal{W} \subset \Rs$. This area takes the shape of a ring sector whose lower radius is $r_1$ and whose upper radius $\Ru$ and arc angle $\rho$ are determined by the type of antenna used. For the type 1 antenna $\Ru = \infty$ and $\rho=2\pi$, as the omnidirectional antenna can receive a signal from any direction with no upper limit on the transmitter distance. For the type 2 antenna the upper radius $\Ru = \Delta \gamma \tan(\omega/2)$ as any transmitters outside of this range will be outside of the main lobe of the UAV antenna, and $\rho = 2\pi$. For the type 3 antenna $\rho= \omega$ as depicted in \Fig{drone_network} and $\Ru$ is given as

\vspace{-4mm}
\begin{align}
\Ru = 
\begin{cases}
\frac{|\Delta \gamma|}{\tan(|\phi_{1}|-\omega/2)} \hspace{-2mm} &\text{if} \hspace{3mm} \omega/2 < |\phi_{1}| < \pi/2 - \omega/2 \\
\frac{|\Delta \gamma|}{\tan(\pi/2 -\omega)} \hspace{-2mm} &\text{if} \hspace{3mm} |\phi_{1}| > \pi/2 - \omega/2 \\
\infty &\text{otherwise}
\end{cases}
\end{align}

\noindent
with $|.|$ denoting absolute value. For the case where $\omega\geq \pi/2$, the major radius $\Ru$ will always be infinite. We denote the set of GSs that fall inside the area $\mathcal{W}$ as $\Phi_{\mathcal{W}} = \{x \in \Phi : x \in \mathcal{W}\}$. The GSs in the set $\Phi_{\mathcal{W}}$ are capable of causing interference to the reference UAV, as their signals may be received by the UAV with non-zero gain. Note that $\Phi_{\mathcal{W}}$ is a PPP with the same intensity $\lambda$.

We assume that the GSs are equipped with tri-sector antennas similar to those already in use in terrestrial BSs, as this allows them to serve UAVs in any horizontal direction. For tractability we model the horizontal antenna gain $\mu_{h}$ of these antennas as having a constant value. The antennas are tilted up towards the sky at an uptilt angle $\phi_T$: the antenna gain from the GS $i$ $\mu(\phi_{i})$ will therefore be a function of the vertical angle of the UAV to $i$, we adopt the 3GPP model of the antenna gain given in Table A.2.1.1-2 in \cite{3GPP_2010}.

The wireless channels between the reference UAV and the GSs will be affected by buildings, which form obstacles and break LOS links. We adopt the model in \cite{ITUR_2012}, which defines an urban environment as a collection of buildings arranged in a square grid. There are $\beta$ buildings per square kilometer, the fraction of area occupied by buildings to the total area is $\delta$, and each building has a height which is a Rayleigh-distributed random variable with scale parameter $\kappa$. The probability of the reference UAV having LOS to the GS $i$ is denoted as $\Pd_{\los}(r_i)$, it is a function of the horizontal distance and is given in Section 2.1.5 of \cite{ITUR_2012}.

We can express the SINR at the reference UAV as $\\p H_{t_{1}} \eta\mu(\phi_{1}) c (r_{1}^2+\Delta \gamma^2)^{-\alpha_{t_{1}}/2}/(I_{\los} + I_{\nlos}+\sigma^2)$, where $p$ is GS transmit power, $H_{t_1}$ is the random multipath fading component, $\alpha_{t_{1}}$ is the pathloss exponent, $t_{1} \in \{\los,\nlos\}$ is an indicator variable which denotes whether the UAV has LOS or NLOS to its serving GS, $\mu(\phi_{1})$ is the serving GS antenna gain, $c$ is the near-field pathloss, $\sigma^2$ is the noise power, and $I_{\los}$ and $I_{\nlos}$ are the aggregate LOS and NLOS interference, respectively. The interferers will belong to the set $\Phi_{\mathcal{W}}$. We partition this set into two subsets which contain the LOS and NLOS interfering GSs, denoted as $\Phi_{\mathcal{W}_{\los}} \subset \Phi_{\mathcal{W}}$ and $\Phi_{\mathcal{W}_{\nlos}} \subset \Phi_{\mathcal{W}}$, respectively. These two sets are inhomogeneous PPPs with intensity functions $\lambda_{\los}(x) = \Pd_{\los}(||x||)\lambda$ and $\lambda_{\nlos}(x) =(1-\Pd_{\los}(||x||))\lambda$. Note that we drop the index $i$ as the GS coordinates have the same distribution irrespective of their index values. The aggregate LOS and NLOS interference is expressed as $I_{\los} = \sum_{x\in\Phi_{\mathcal{W}_{\los}}} p H_{\los} \eta\mu(\phi) c (||x||^2+\Delta \gamma^2)^{-\alpha_{\los}/2}$ and $I_{\nlos} = \sum_{x\in\Phi_{\mathcal{W}_{\nlos}}} p H_{\nlos} \eta\mu(\phi) c (||x||^2+\Delta \gamma^2)^{-\alpha_{\nlos}/2}$, recalling that $\phi = \arctan(\Delta \gamma/||x||)$. 
 
  The probability that the SINR between a UAV and its serving ground station exceeds some threshold $\theta$ is referred to as the backhaul probability of the reference UAV. An analytical expression for this probability 
%when the UAV is connected to the GS network 
is derived in the next section.

%\subsection{Millimeter-wave Backhaul}

%For the millimeter-wave backhaul we assume each GS is equipped with an antenna array that uses beamforming to direct a directional beam towards the UAV to which it provides a backhaul %link. We adopt a similar approach to modelling the GS antenna array as in \cite{Elshaer_2016} and \cite{Andrews_2017}. The GS antenna is modelled as having a single directional beam with a beamwidth of $\omega_G$ and a gain of $\mu_{m}$ inside the main lobe, and a gain of 0 outside. The reference UAV will always experience an antenna gain of $\mu_{m}$ from its serving GS. The beam patterns of the remaining GSs will appear to be pointed in random directions with respect to the reference UAV; as a result each interfering GS will have non-zero antenna gain to the reference UAV with a certain probability $\zeta$. 
\vspace{-2mm}
\section{Analytical Results}
In this section we derive an analytical expression for the probability that the reference UAV will receive a signal from the GS network with an SINR above a threshold $\theta$.
%, thereby meeting the C\&C link requirements. 
We derive the backhaul probability by, first, finding the conditional probability that the UAV receives a signal of SINR above $\theta$ given that the serving GS is located at a distance $r_1$ and that the channel between the UAV and that GS is LOS/NLOS. We then decondition this conditional backhaul probability with respect to the LOS probability of the serving GS as well as its horizontal distance. Given a PPP distribution of GSs, the serving GS horizontal distance random variable $R_1$ is known to be Rayleigh-distributed with scale parameter $1/\sqrt{2\pi\lambda}$.

\vspace{-3mm}
\subsection{Conditional Backhaul Probability}
\vspace{-2mm}
%In this subsection we describe how the coverage probability for the UAV network is obtained as a function of the network design and wireless channel parameters.
\noindent
%Considering Nakagami-$m$ fading, the conditional coverage probability $\Pd(\sinr\geq \theta |R_1=r_1)$ is obtained following (21) in \cite{Chetlur_2017} as

%\begin{equation}
 %\Pd(\sinr\geq \theta |R_1=r_1) =  \\
% \sum\limits_{n=0}^{m_{t_1}-1}\frac{s_{t_1}^n}{n!} (-1)^n \frac{d^n %\Lc_{I}((p\eta)^{-1}s_{t_1})}{ds_{t_1}^n},
%\end{equation}

%\noindent
%where $s_{t_1}= m_{t_1}(\eta_{GS}(\phi_1)^{-1})\theta(r_1^2+\Delta\gamma^2)^{\alpha_{t_1}/2}$, $m_{t_1}$ is the Nakagami-$m$ fading term and $\Lc_{I}$ denotes the Laplace transform of the total interference. The LOS and NLOS interferers are distributed independently of one another; the proof of this is similar to the proof in \cite{Haenggi_2013} and is omitted here. Due to this, the Laplace transform above can be separated into a product of the Laplace transforms of the aggregate LOS and aggregate NLOS interference, along with the introduction of the noise-related term. This allows us to express the conditional coverage probability for an LOS serving UAV $\Pd(\sinr\geq \theta |R_1=r_1,t_1 = \textit{l})$ as 

The expression for the backhaul probability, given serving GS distance $r_1$ and an LOS channel to the serving GS $\Pd(\sinr\geq \theta |R_1=r_1,t_1 = \los)$, follows the derivation provided by us in \cite{Galkin_2017} as
\vspace{-3mm}
\begin{align}
&\sum\limits_{k=0}^{m_{\los}-1}\frac{s_{\los}^k}{k!} (-1)^k  \sum_{i_{\los}+i_{\nlos}+i_{\sigma}=k}\frac{k!}{i_{\los}!i_{\nlos}!i_{\sigma}!} \nonumber \\
 %&\cdot(-(p\etac)^{-1}\sigma^2)^{i_{\sigma}}\exp(-(p\etac)^{-1}s_{\los}\sigma^2) \nonumber \\
 &\frac{d^{i_{\los}} \Lc_{I_{\los}}((p \eta c)^{-1}s_{\los})}{ds_{\los}^{i_{\los}}} \frac{d^{i_{\nlos}}\Lc_{I_{\nlos}}((p  \eta c)^{-1}s_{\los})}{ds_{\los}^{i_{\nlos}}} \frac{d^{i_{\sigma}}\exp(-(p\eta c)^{-1}s_{\los}\sigma^2)}{ds_{\los}^{i_{\sigma}}},
\label{eq:condProb3}
\vspace{-3mm}
\end{align}

\noindent
where $s_{\los}= m_{\los}\theta \mu(\phi_1)^{-1}(r_1^2+\Delta\gamma^2)^{\alpha_{\los}/2}$, $m_{\los}$ is the Nakagami-$m$ fading term for a LOS channel, $\Lc_{I_{\los}}$ and $\Lc_{I_{\nlos}}$ are the Laplace transforms of the aggregate LOS and NLOS interference, respectively, and the second sum is over all the combinations of non-negative integers $i_{\los},i_{\nlos}$ and $i_{\sigma}$ that add up to $k$. The conditional backhaul probability given an NLOS serving GS $\Pd(\sinr\geq \theta |R_1=r_1,t_1 = \nlos)$ is calculated as in \Eq{condProb3} with $m_{\nlos}$, $\alpha_{\nlos}$ and $s_{\nlos}$ replacing $m_{\los}$, $\alpha_{\los}$ and $s_{\los}$. 

\subsection{Laplace Transform of Aggregate Interference}
%\textbf{LTE backhaul}
 Following the derivation process of Eq. (8) in \cite{Galkin_2018} the Laplace transform of the aggregate LOS interference $\Lc_{I_{\los}}((p\eta c)^{-1}s_{\los})$ given an LOS serving GS is expressed as
% \vspace{-2mm}
\begin{align}
\vspace{-5mm}
&\exp\hspace{-1mm}\Bigg(\hspace{-1mm}-\hspace{-1mm}\lambda \rho  \int\limits_{r_1}^{\Ru}\Bigg(\hspace{-1mm}1- \hspace{-1mm}\left(\frac{m_{\los}}{g(r,s_{\los},\alpha_{\los})+m_{\los}}\right)^{m_{\los}}\hspace{-1mm}\Bigg) \Pd_{\los}(r)r \dr r \hspace{-1mm}\Bigg) \nonumber \\
\label{eq:laplace}
\end{align}

\vspace{-5mm}
\noindent
where $g(r,s_{\los},\alpha_{\los}) = s_{\los}\mu(\phi)(r^2+\Delta\gamma^2)^{-\alpha_{\los}/2} $.

\noindent
 Note that the expression for the Laplace transform for the NLOS interferers $\Lc_{I_\nlos}((p \eta c)^{-1}s_\los)$ is obtained by simply substituting $\Pd_{\los}(r)$ with $(1-\Pd_{\los}(r))$, $m_\los$ with $m_\nlos$ and $g(r,s_\los,\alpha_\los)$ with $g(r,s_\los,\alpha_\nlos)$ in \Eq{laplace}. The above integration is for the case when the serving GS is LOS; if the serving GS is NLOS we substitute $s_\los$ with $s_\nlos$ as defined in the previous subsection.

\vspace{-3mm}
\subsection{Backhaul Probability}

%\begin{proposition}
Using the derivations provided in the previous subsections the overall backhaul probability can be expressed as%To obtain the overall backhaul probability for the reference UAV in the network we decondition the conditional backhaul probability as defined in the previous subsection with respect to the indicator variable $t_1$ (we use the LOS probability function in \Eq{LOS}). We then decondition with respect to the horizontal distance random variable $R_1$.
%\end{proposition}
% \vspace{-3mm}
\begin{align}
&\Pd(\sinr\geq \theta) = 
\hspace{-1mm}\int\limits_{0}^{\infty}\hspace{-1mm}\bigg(\Pd(\sinr\geq \theta |R_1=r_1,t_1 = \los)\Pd_{\los}(r_1) \nonumber
\end{align}
\vspace{-7mm}
\begin{align}
\vspace{-10mm}
&+\Pd(\sinr\geq \theta |R_1=r_1,t_1 = \nlos)(1-\Pd_{\los}(r_1))\bigg)f_{R_1}(r_1) \dr r_1 .
 \label{eq:pcov_final}
\end{align}

\vspace{-6mm}

 \section{Numerical Results}

In this section we use our model to evaluate the performance of the GS network, and compare it to the performance that can be achieved from the UAVs connecting to an existing terrestrial BS network. We simulate a PPP distribution of terrestrial BSs with a density of \unit[5]{$/km^2$} and downtilted antennas \cite{3GPP_2010} over multiple MC trials. We set $\omega$ to \unit[165]{deg} for the fixed directional antenna, \unit[60]{deg} for the steerable directional antenna, $\alpha_{\los}$ and $\alpha_{\nlos}$ are set to 2.1 and 4, respectively, $m_{\los}$ and $m_{\nlos}$ are both set to 1, transmit power $p$ is \unit[40]{W}, $\mu_{h}$ is set to \unit[-5]{dB}, values of $c$ and $\sigma^2$ are \unit[-38.4]{dB} and \unit[$8\cdot10^{-13}$]{W}, respectively \cite{Elshaer_2016}, $\theta$ is set to \unit[10]{dB}, $\phi_T$ is taken as $\arctan(\Delta\gamma/\Ed [R_1] )$, $\gamma_{G}$ is set to \unit[30]{m}, the LOS model parameters $\beta$, $\delta$, $\kappa$ are taken as \unit[300]{$/\text{km}^2$}, 0.5 and \unit[20]{m}, respectively.
 %Unless stated otherwise the parameters used for the numerical results are taken from Table \ref{tab:table}.

\begin{figure}[b!]
\vspace{-5mm}
\centering
	\includegraphics[width=.45\textwidth]{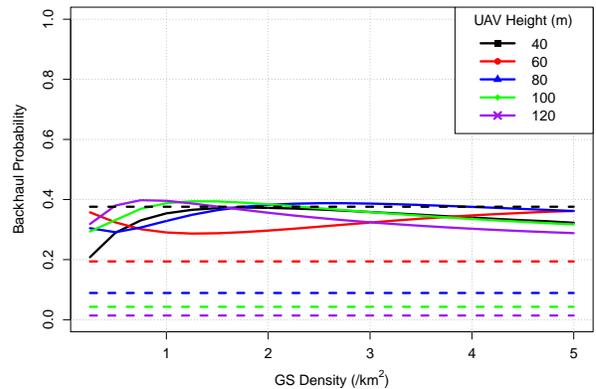}
    \vspace{-6mm}
	\caption{
	Coverage probability as a function of the GS density and UAV height for the omnidirectional UAV antenna case. Solid lines denote the GS network, dashed lines denote the terrestrial BS network.
	}
	\label{fig:DensityType1}
	\vspace{-3mm}
\end{figure}

In \Fig{DensityType1} we compare the coverage probability achieved by the dedicated GS network against a typical terrestrial BS network, when UAVs are equipped with omnidirectional antennas. We can see that the performance for both networks is relatively poor.
%, although as expected the dedicated GS network outperforms the use of existing base stations.
%, when comparing to the results for the other antenna types. 
In both cases, omnidirectional UAV antennas receive a large amount of interference, 
and the greater the height at which the UAV operates, the greater the likelihood of LOS interferers.
%at greater heights the UAV is exposed to even more LOS interferers which can severely affect the network performance. 
We observe that the dedicated GS network provides a superior signal quality compared to the terrestrial network, particularly at greater UAV heights. The downtilted antennas of the terrestrial BSs mean that at large heights the UAV receives side-lobe signals from its serving BS; combined with the greater interference strength, this causes significant reduction of signal quality. The GS network is able to provide a signal at full strength with its antennas which are tilted towards the sky, and as a result the achievable coverage probability remains relatively stable for the different UAV heights. Note that the GS network only requires a fraction of the density of the terrestrial network to achieve equivalent or better performance.

\begin{figure}[t!]
\vspace{-5mm}
\centering
	\includegraphics[width=.45\textwidth]{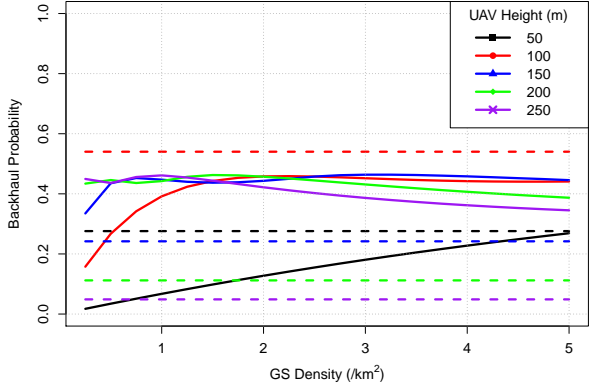}
    \vspace{-6mm}
	\caption{
Coverage probability as a function of GS density and UAV height for the downtilted directional UAV antenna case. Solid lines denote the GS network, dashed lines denote the terrestrial BS network.
	}
	\label{fig:DensityType2}
%	\vspace{-3mm}
\end{figure}

In \Fig{DensityType2} we consider a UAV equipped with a downtilted directional antenna. The antenna directionality means that the UAV will only receive signals from devices within a certain horizontal distance of its location, with the UAV height determining this distance. When connecting to the BS network, a greater distance means that the UAV can connect to BSs further away which provide a stronger signal due to better BS antenna alignment \cite{MahdiAzari_20172}; however, it also means that the UAV will receive interference from more devices. The dedicated GS network with uptilted antennas can outperform the terrestrial BS service, but only when the UAV operates above a certain height. The decision to deploy a dedicated GS network for UAV service is therefore based on the specific height range the target UAVs will operate at.

\begin{figure}[t!]
\vspace{-5mm}
\centering
	\includegraphics[width=.45\textwidth]{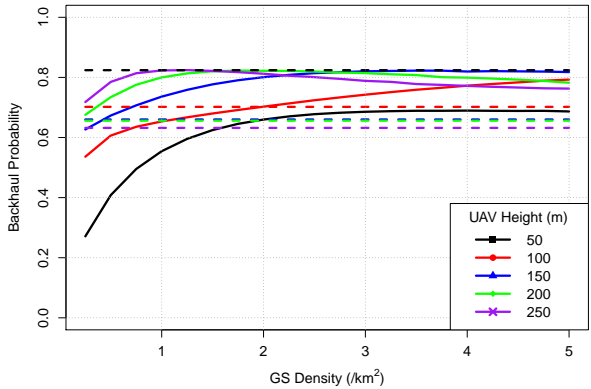}
    \vspace{-6mm}
	\caption{
	Coverage probability as a function of GS density and UAV height for the steerable directional UAV antenna case. Solid lines denote the GS network, dashed lines denote the terrestrial BS network.
	}
	\label{fig:DensityType3}
	\vspace{-3mm}
\end{figure}

In \Fig{DensityType3} we show the network performance when the UAV is equipped with a steerable antenna, which it points at its serving GS or BS. Similar to the results in \Fig{DensityType2}, the GS network outperforms the BS network when the UAVs operate at sufficiently large heights. Note that the coverage probability for the terrestrial BS case is improved over the previous two antenna cases, with the steerable UAV antenna being able to reduce the amount of received interference. The results show that equipping the UAV with a high quality antenna may provide a wireless signal of sufficient quality even through an unmodified terrestrial BS network.

	\vspace{-5mm}

\section{Discussion}
\vspace{-2mm}
In this paper we evaluated the performance of a dedicated  GS network to support high quality links to UAVs, given different types of UAV antennas. Using stochastic geometry and simulations we demonstrated that a dedicated GS network can give superior performance to a UAV when compared to terrestrial BS networks, unless the UAV is operating at a sufficiently low height, or is equipped with a high quality antenna which can be intelligently steered towards a transmitter. 

Given the wide variety of UAV applications emerging on the market today, with different data requirements, operating heights, flight behaviours and antenna configurations, the demand for connectivity to these types of vehicles will continue to grow.
A network operator must carefully consider these use cases, avoiding overprovisioning the wireless network by deploying dedicated infrastructure for serving UAVs that do not require it, while considering the deployment of dedicated ground stations in locations where high demand for UAV connectivity justifies this investment.
%forcing UAVs to rely on existing infrastructure which is wholly unsuitable for the UAV's unique operational requirements. 
From the perspective of the UAV operator, the results also demonstrate the importance of being aware of the underlying communications infrastructure and its limitations. Given the mobility of UAVs and their flexibility, many connectivity issues may be resolved by the UAV operator adjusting the UAV behaviour around the limitations of the infrastructure, rather than the network operator upgrading the infrastructure around the needs of the UAVs. 

 \vspace{-6mm}
\section*{Acknowledgements}
 \vspace{-2mm}
This material is based upon works supported by the Science Foundation
Ireland under Grants No. 14/US/I3110 and 13/RC/2077 (CONNECT).

\ifCLASSOPTIONcaptionsoff
  \newpage
\fi



 \vspace{-5mm}
\bibliographystyle{./bib/IEEEtran}
\bibliography{./bib/IEEEabrv,./bib/IEEEfull}
\end{document}